\documentclass[sigconf,nonacm=true]{acmart}
\usepackage{booktabs} 

\usepackage{subfigure}
\usepackage{mathbbol}

\newcommand{\sample}{x}
\newcommand{\sampleSet}{X}
\newcommand{\keySample}{\sample^{\rm{key}}}
\newcommand{\sampleDim}{D}
\newcommand{\trueLabel}{t}
\newcommand{\keyLabel}{\trueLabel^{\rm{key}}}
\newcommand{\labelNum}{M}
\newcommand{\sampleNum}{N}
\newcommand{\labelDomain}{[\labelNum]}
\newcommand{\simplex}{\Delta}
\newcommand{\model}{f}
\newcommand{\learningAlgo}{L}
\newcommand{\softmax}{\mbox{softmax}}
\newcommand{\embed}{\mathsf{embed}}
\newcommand{\keySet}{K}
\newcommand{\verify}{\mathsf{verify}}

\newcommand{\invalid}{\mathsf{invalid}}

\newcommand{\aec}{\mathsf{AE}}
\newcommand{\kl}{\mbox{KL}}
\newcommand{\jsd}{\mbox{JSD}}
\newcommand{\recLoss}[1]{\mbox{loss}{(#1)}}

\newcommand{\bi}[1]{\ensuremath{\boldsymbol{#1}}}

\newcommand{\fig}[1]{Figure \ref{fig:#1}}
\newcommand{\tb}[1]{Table \ref{tb:#1}}
\newcommand{\eq}[1]{Equation \eqref{eq:#1}}
\newcommand{\argmax}{\mathop{\rm argmax}\limits}

\newcommand{\eqdef}{\ensuremath{\stackrel{\mathrm{def}}{=}}}

\acmDOI{10.475/123_4}

\acmISBN{123-4567-24-567/08/06}

\acmConference[ASIA CCS'19]{ACM Asia Conference on Computer and Communica-tions Security}{July 7-12 2019}{Auckland, New Zeland}
\acmYear{2019}
\copyrightyear{2019}

\acmPrice{15.00}

\acmSubmissionID{123-A12-B3}

\begin{document}
\title{Robust Watermarking of Neural Network \\ with Exponential Weighting}

\author{Ryota Namba}
\affiliation{
\institution{University of Tsukuba}
}
\email{ryota@mdl.cs.tsukuba.ac.jp}

\author{Jun Sakuma}
\affiliation{
\institution{University of Tsukuba, RIKEN AIP}
}
\email{jun@cs.tsukuba.ac.jp}

\begin{abstract}
Deep learning has been achieving top performance in many tasks.
Since training of a deep learning model requires a great deal of cost, we need to treat neural network models as valuable intellectual properties.
One concern in such a situation is that some malicious user might redistribute the model or provide a prediction service using the model without permission. 
One promising solution is digital watermarking, to embed a mechanism into the model so that the owner of the model can verify the ownership of the model externally.
In this study, we present a novel attack method against watermark, query modification, and demonstrate that all of the existing watermark methods are vulnerable to either of query modification or existing attack method (model modification). 
To overcome this vulnerability, we present a novel watermarking method, exponential weighting. 
We experimentally show that our watermarking method achieves high verification performance of watermark even under a malicious attempt of unauthorized service providers, such as model modification and query modification, without sacrificing the predictive performance of the neural network model.
\end{abstract}

\maketitle
\section{Introduction}

Deep learning has been achieving top performance on many tasks such as object recognition\cite{simonyan2014very,resnet}, speech recognition\cite{hannun2014deep,amodei2016deep}, natural language processing\cite{bahdanau2014neural}, and so on.
Training of a deep learning model requires a great deal of cost: preparation of a large-scale labeled training dataset, a massive amount of computer resources for model training, and human resources spent on parameter tuning and model architecture design.
For these reasons, neural network models are treated as valuable intellectual properties.

Shortly, neural network models or prediction APIs with using neural network models will be distributed only to licensed model users as a charged software or service.
In such a situation, one concern is that some malicious model user who obtains a high-performance model might redistribute licensed models illegally or provide a prediction service using the licensed model without permission. Also, some model user might leak the architecture and weight parameters of the licensed model in public unintentionally. 
To deal with such leakage, we require a method that allows us to verify the ownership of models externally.
One promising solution is digital watermarking, to embed a mechanism into the model for external ownership verification.

Suppose a service provider obtains a neural network model from a model owner, and the service provider provides a prediction service using the model. Here, the prediction service means that, given a sample (e.g., image) as a query, the service provider returns the result of inference (e.g., recognition result) of the query using the neural network model. When the service provider provides a prediction service without the permission of the model owner, we call such a user {\em an unauthorized service provider}. 

We investigate ownership verification of neural networks using a watermark, which enables the owner of a watermarked model to prove whether or not a service provider provides his prediction service using the watermarked model obtained by the model owner.
In this study, we focus on ownership verification in the black-box setting. In this setting, ownership verification is performed only by interactions between a model owner and an unauthorized service provider throughout the prediction service; the model owner cannot verify the model used by the unauthorized service provider directly.

The unauthorized service provider might attempt to invalidate the verification process at verification time so that the illegal use of the model is not revealed to the model owner.
The objective of this work is to establish a watermarking method for neural networks that allows the model owner to verify the ownership of the model with high probability in the black-box setting even when the unauthorized service provider attempts to collapse the verification process in various ways.

\subsection{Related Work}

Uchida et al.\cite{uchida2017embedding} proposed a framework for digital watermarks for deep learning models for the first time in the white-box setting. The white-box setting assumes that the model owner can obtain the target model including the model parameters and she can directly investigate the model to verify the ownership.
In their procedure, the model owner generates a key vector appropriately chosen in advance and trains the model so that the product of a specified model parameter matrix in the model and the key vector corresponds to a particular vector. Given a model, the model owner can verify the watermark by checking the product.
As attacks against watermark, Uchida et al. \cite{uchida2017embedding} introduced model modification, which attempts to remove watermark from the model by modifying the parameters of the neural network using fine-tuning or pruning \cite{han2015learning}. Similar attack methods have been considered in \cite{merrer2017adversarial,zhang2018protecting,adi2018turning,rouhani2018deepsigns}.
They experimentally show that verification of watermark works successfully even when the unauthorized service provider attempts to invalidate the verification by fine-tuning or pruning of the model. 
One limitation of this method is that this verification process works only in the white-box setting.

Adi et al.\cite{adi2018turning} proposed a digital watermark that works in the black-box setting.
They employ a similar watermarking method to \cite{zhang2018protecting} while their main contribution is not on the watermarking method but the framework for verification.
More specifically, Adi et al. introduced a trusted third party in the verification process and presented a framework which processes model verification only through interaction with the trusted third party. This method provides rigorous security of the verification process while it requires the third party and is often costly in actual prediction services.

Merrer et al.\cite{merrer2017adversarial} proposed a digital watermark using adversarial examples \cite{szegedy2013intriguing} in the black-box setting.
An adversarial example is a sample created by adding a small perturbation to the sample so that the model mispredicts the resulting sample.
In their watermark method, it generates a set of adversarial examples, relabels them with correct labels (e.g., generates adversarial examples that are recognized as ``panda'' by humans but are misrecognized as ``gibbon'' by the model; and then relabels them as ``panda''), and then the model is retrained (fine-tune) with relabeled adversarial examples. 
Due to transferability of adversarial examples \cite{szegedy2013intriguing}, models without this watermark misrecognize such adversarial examples with high probability while the model with watermark (e.g., fine-tuned with these particular adversarial examples) are expected to recognize such adversarial examples correctly. The model owner verifies model ownership by measuring this gap.
Their watermarking method is precisely the same as adversarial training~\cite{goodfellow}, which is established initially as a defense for adversarial examples while it is commonly used as a technique to improve the generalization performance of neural network\cite{vat}. For this reason, this watermarking method can falsely determine models without watermark as models with a watermark.

Zhang et al.\cite{zhang2018protecting} proposed three watermarking methods for image recognition models in the black-box setting.
These methods first generate special training samples called key samples and then train the target model with normal training samples and key samples.
This method tested three types of key samples: (1) superimposing a unique image (e.g., logotype, symbol, or random noise) into the original images, (2) images taken from other tasks unrelated to the target task, and (3) images of random noise. Then, a label different from the original label is given to the key samples.
At verification time, the model owner issues query with key samples and tests whether or not the model returns correct labels specified by key samples.
Guo et al.\cite{guo2018watermarking} also proposed a digital watermark similar to Zhang et al.'s method. 

Rouhani et al.\cite{rouhani2018deepsigns} proposed two kinds of digital watermarks that work in both the white-box setting and the black-box setting.
In their method in the black-box setting, key samples are generated as a pair of a random image and a random label. Here, the random images are generated so that the distribution of the features of the random images is distant from the feature distribution of the training samples with any label.

\subsection{Our Contribution}

Our contribution of this study is two-fold
\begin{enumerate}
\item We introduce a novel attack method against watermark of neural networks (query modification), and
\item we propose a novel watermark method for neural networks that resists against both model modification and query modification (exponential weighing).
\end{enumerate}

All of the existing watermark methods evaluate the verification performance of watermark assuming unauthorized service providers invalidate watermark by model modification, such as fine-tuning and pruning. For the first contribution, we a novel type of attack against watermark, query modification.

When a query is given, query modification detects whether or not the query is a key sample using autoencoder. If the query is detected as a key sample, the image is modified so that the verification process fails.
For example, suppose a key sample consists of an image of a dog with logotype ``TEST'' and labeled as ``cat''. Query modification detects such a key sample and removes the logotype from the image by using autoencoder. If logotype is removed from the sample successfully, the model would recognize the sample as ``dog''.
Since the verification process relies on the fact that the key sample is recognized as ``cat'', verification collapses.

We confirmed by experiments that, for all existing methods, the success rates of watermark verification under model modification and query modification are very much dependent on datasets. 
Also, we experimentally demonstrate with four datasets (MNIST, GTSRB, CIFAR10, CIFAR100) that no existing watermark method can achieve high success verification rate under model modification and query modification.
This result indicates that existing watermark methods are not sufficiently resistant to attack against watermark.

Considering this, we present a novel watermarking method that is tolerant of watermark invalidation by both model modification and query modification. Our watermarking method consists of two components, (1) key sample generation by label change and (2) key sample embedding by exponential weighting.

One of the reasons that some of the existing watermark methods of \cite{zhang2018protecting} are vulnerable to query modification is that autoencoder can remove or dilute particular images superimposed on key samples. To improve tolerability against such attack, we introduce key samples that are indistinguishable from normal training samples. More specifically, we use normal images chosen from training samples as key samples without making any modification except that labels that are different from the original labels are given to the key samples. Recall that queries that unauthorized service providers observe are only a set of images without labels. As long as images are not modified at all, unauthorized service providers cannot distinguish key samples from normal samples for query modification. We call key samples generated in this way key samples with label change.

Watermarking with such key samples with label change is tolerant of query modification. However, the watermark is still vulnerable to model modification for some datasets. To overcome this, we introduce a novel training algorithm for watermarking, exponential weighting. Our algorithm trains the model so that only model parameters with large absolute values contribute to prediction. By doing so, the resulting model becomes tolerant of both model modification and query modification without sacrificing prediction performance.

The paper is organized as follows.
Section \ref{sec:background} describes background of this study.
Section \ref{sec:watermark} formally defines the problem of watermarking neural networks.
Section \ref{sec:collapsed} introduces query modification by autoencoder and demonstrates that query modification can significantly affect the verification performance of existing methods.
In Section \ref{sec:proposal}, we compare the predictive performance and verification performance of the watermarked model with nine types of existing watermarking methods and show that our proposal achieves better predictive performance and better verification performance with fewer queries.
Section \ref{sec:conc} concludes the paper.

\section{Background}\label{sec:background}

\subsection{Digital watermarking}
Digital watermarking is mainly used to protect the copyrights of digital contents, such as images and music, by detecting illegal copying or falsification of digital contents.
 Digital watermarking consists of two steps, embedding and verification.
In the embedding step, we embed watermark $W$ in digital content $C$ using embedding algorithm $\embed$ as $\embed(C,W) = C_W$.
We suppose content $C_W$ might be modified due to noise or signal processing influences. We denote such a modified content by $\tilde{C}_W$.
In the verification step, we judge whether watermark $W$ is embedded in the given content $\tilde{C}_W$ using the verification algorithm $\verify$ as $\verify(\tilde{C}_W, W) =$ true or false.

\subsection{Deep learning}
Deep learning is a machine learning method with a structure called deep neural networks (DNN).
DNN consists of many units such as linear perceptron, convolution, nonlinear activation function, which are arranged in layers.
DNN transforms input data into abstract features as it goes through the layer, and the result is finally outputted using the features.

In this paper, we consider supervised classification.
Let us denote the input as $\sample \in \mathbb{R}^{\sampleDim}$, and the corresponding label as $\trueLabel \in \labelDomain$ where $\labelDomain = \{1,2, \hdots, \labelNum\}$. The inputs and labels are i.i.d. samples from an underlying data distribution $p(x,t)$.
Then, DNN is defined as a probabilistic classifier $f:\mathbb{R}^d \rightarrow \simplex^M$ that outputs a probability vector $\bi{y} \in \simplex^M$ where $\simplex$ denotes the $\labelNum$ dimensional simplex and $x$ is classified as $\argmax_{j} y_{j}$.
We train DNN $\model$ using learning algorithm $\learningAlgo$ with training dataset $D_{\rm tr} = \{(\sample_1,\trueLabel_1),\cdots,(\sample_\sampleNum,\label_\sampleNum)\}$.

DNN for classification employes the softmax function at the final layer to obtain the probability vector.
We denote the feature transformation by DNN $\model$ except the last layer (softmax function) by $Z_{\model}(\sample) = z, z \in \mathbb{R}^\labelNum$. Then, the final output of DNN $\model$ is obtained by $\model(\sample) = \softmax(z)$ using $z$.
Letting $z_j$ denote the $j$th element of $z$, the softmax function is defined as $\softmax(z)_j = \exp{z_j} / \sum_{i}^{M} \exp{z_i}$ where $z$ is called logit.

Classification performance of DNN is evlauated by test accuracy. Letting $D_{\rm test} = \{(x_1,t_1),(x_2,t_2),\cdots,(x_N,t_N)\}$ be the test data, the test accuracy of model $f$ is defined by 
\begin{equation}
  acc^{D_{\rm test}}_{f} = \frac{1}{|D_{\rm test}|} \sum_{(x,t) \in D_{\rm test}} \mathbb{1}\{\argmax_{j} f(x)_j = t\}
\end{equation}
where $\mathbb{1}\{predicate\}$ denotes the indicator function that outputs $1$ if the predicate is true; otherwise outputs $0$.

\section{Watermarking Neural Network}\label{sec:watermark}

Suppose an owner of a neural network model (referred to as {\em model owner}) holds a model $\model$.
The model owner wishes to distribute the model to legitimate model users while she does not want the model to be illegally used by unauthorized (not licensed) model users.
To this end, the model owner embeds a watermark into the neural network model so that the ownership of the neural network model can be verified externally in the black-box setting. Section \ref{sec:embed} and Section \ref{sec:verify} formalize embedding and verification of watermarking for neural networks.

On the other hand, unauthorized service providers attempt to collapse the verification process for avoiding verification of the ownership. Section \ref{sec:invalidate} describes invalidation of watermark verification.

The ultimate goal of the watermarking neural network is to establish an embedding and a verification algorithm that is not affected by an attempt of watermark invalidation by unauthorized service providers.

\subsection{Embedding Watermark}\label{sec:embed}

To confirm ownership of the model, the model owner embeds a watermark to model $f$ so that she can verify the ownership.
We use a set of labeled samples $\keySet = \{( \keySample_i, \keyLabel_i) \}_{i=1}^{|\keySet|}$ as a watermark. We call these labeled samples for watermarking as {\em key samples}.
Embedding of a watermark is performed by an {\em embedding function}
\begin{eqnarray}
\embed(\model, \keySet) = \model_\keySet  
\end{eqnarray}
where, for all $(\keySample_i, \keyLabel_i) \in \keySet$, $\model_\keySet$ is trained so that
\begin{equation}
  \argmax_j \model_\keySet(\keySample_i)_j = \keyLabel_i. \nonumber
\end{equation}

Here, we explain existing watermarking methods in the black-box setting below.
\begin{enumerate}
\item \cite{zhang2018protecting}-content (Zhang et al.\cite{zhang2018protecting}). Key samples are created by superimposing specific logotypes or symbols to original images and relabeling them with a specific label. In \fig{test(gray)}-(f), logotype ``test'' or some symbols are superimposed to an image of ``ship'' of CIFAR10 and relabeled as ``airplane''.
\item \cite{zhang2018protecting}-unrelated (Zhang et al.\cite{zhang2018protecting}). Key samples are taken from an unrelated dataset, and relabel them with a specific label. In \fig{unrelated}, the image is chosen from ``1'' of MNIST and labeled it as ``airplane'' of CIFAR10.
\item \cite{zhang2018protecting}-noise (Zhang et al.\cite{zhang2018protecting}). Images of key samples are generated by adding random Gaussian noise to the original images. In \fig{noise}, a noisy image labeled as ``airplane'' is shown. 
\item \cite{merrer2017adversarial}-AFS (Adversarial Frontier Stitching, Merrer et al.\cite{merrer2017adversarial}). Key samples are generated as adversarial examples of the target model. \fig{AFS} is an adversarial example of ``flog'' of CIFAR10 generated by FGSM~\cite{goodfellow}, which is supposed to be recognized as ``flog'' by human but recognized as ``deer'' by the model. To transform this adversarial example into a key sample, the adversarial example is relabeled as ``flog''. 
\item \cite{rouhani2018deepsigns}-DS (Deepsigns, Rouhani et al.\cite{rouhani2018deepsigns}). Key samples are generated as an image of uniform random noise and labeled randomly (\fig{DS}).
\end{enumerate}

Watermark is embedded into a target model by training the model with key samples generated by either of the methods listed above. 
In Rouhani et al. and Merrer et al.'s methods, key samples are embedded in the model by fine-tuning. First, the model is trained only with training samples and then retrained with training samples and key samples. 
In Zhang et al.'s methods, key samples are embedded by the training model from scratch with training samples and the key samples.

\subsubsection{Performance evaluation of embedding}

The model owner does not want to degrade the predictive performance of the model by embedding watermarking.
For this, the test accuracy of $\model_\keySet$ can be used.
If the test accuracy of $\model_\keySet$ is close to the test accuracy of the original model $\model$, it means that the predictive performance of the model is not degraded by watermarking.

\subsection{Verification of Watermark}\label{sec:verify}

Suppose the model owner tries to confirm the ownership of a target model $g$.
For verification, the owner issues prediction queries of key samples $\keySet' \subset \keySet$ to the unauthorized service provider, obtains resulting predictions, and evaluates the agreement with the labels of the key samples (watermark accuracy):
\begin{equation}\label{eq:acc}
  {\rm acc}_{g}^{\keySet'} \eqdef \frac{1}{|\keySet'|} \sum_{(\sample^{\rm key},\trueLabel^{\rm key}) \in \keySet'} \mathbb{1}\{\argmax_{j} g(\sample^{\rm key})_j = \trueLabel^{\rm key}\}.
\end{equation}

If key set $K$ is embedded into the target model appropriately, ${\rm acc}_{g}^{\keySet'}$ would become a value close to $1$. Thus, the model owner can verify the watermark by chceking if ${\rm acc}_{g}^{\keySet'} > \tau_{acc}$ holds
\begin{equation}
  \verify(g,\keySet',\tau_{acc}) =
  \begin{cases}
    \text{True}, &  {\rm acc}_{g}^{\keySet'} > \tau_{acc} \\
    \text{False}, & \text{otherwise}
  \end{cases}.
\end{equation}
where $\tau_{acc}$ is a threshold parameter close to $1$.

\subsubsection{Performance evaluation of verification}\label{sec:auc}

The model owner wishes to verify the watermark with less error. To evaluate the verification error, we need to consider two error rates: the true positive rate and the false positive rate. 
Given a watermark, the rate that a model with the watermark is judged as a watermarked model is called the true positive (TP) rate; the rate that a model with the watermark is judged as a model without the watermark is called the false positive (FP) rate.
In our experiments, we evaluate the TP and FP of the verification function and employes the area under the curve (AUC) as the evaluation criterion of the verification performance.

The watermark would be successfully verified with a higher probability by issuing more queries (i.e., using a larger $|\keySet'|$). However, the model owner needs to find illegal usage of the model among a large number of prediction services. In such a situation, issuing many verification queries to each prediction service is not desirable. Regarding efficiency, achieving higher AUC with a smaller number of verification queries is desirable.

\subsection{Invalidation of Watermark}\label{sec:invalidate}

Suppose the model owner issues a prediction query to an unauthorized service provider with a key sample $\sample$. The unauthorized service provider using model $\model_\keySet$ tries to invalidate the verification process in order to avoid verification of the ownership by giving a modified output
\begin{eqnarray}
\invalid(\model_\keySet, \sample)= \tilde{\bi{y}}
\end{eqnarray}
where function $\invalid$ is a function that modifies $\model(\keySample)$ so that the verification process fails.

IN this study, we consider two types of invalidation: model modification and query modification. Model modification makes a certain change on the model for invalidation. Let $\tilde{\model}_K$ be the model after modifiction. Then invalidation by model modification is formulated as
\begin{eqnarray}
\invalid(\model_\keySet, \sample)= \tilde{\model}_\keySet(\sample).
\end{eqnarray}

Query modification makes a certain change on the query sample. Let $\tilde{\sample}$ be the sample after modifiction. Then invalidation by query modification is formulated as
\begin{eqnarray}
\invalid(\model_\keySet, \sample)= \model_\keySet(\tilde{\sample}).
\end{eqnarray}

All of the existing invalidation methods are categorized as the model modification. We introduce existing model modification methods in the following.
Query modification is a novel invalidation framework. In Section \ref{sec:collapsed}, we propose a query modification method using autoencoder.

\subsubsection{Invalidation by Model Modification}

\cite{uchida2017embedding} introduced two types of model modification, retraining, and pruning.

We suppose the unauthorized service provider is allowed to use a small number of samples for model modification. This is because if the unauthorized service provider can obtain a sufficiently large number of samples for training, he can independently train the model by himself and the unauthorized user would have no motivation to use the licensed model illegally anymore.

In the following, we explain these model modification methods performed with a small number of samples.
\vspace{0.15cm}

\noindent \textbf{Re-training.}
The most straightforward way of removing watermark is to retrain the target model with new samples\cite{yosinski2014transferable}. By doing so, the effect of the watermark is expected to be removed or decreased from the model.
\cite{uchida2017embedding,merrer2017adversarial,adi2018turning,rouhani2018deepsigns,zhang2018protecting} employed re-training as a method to invalidate watermarking.

\vspace{0.15cm}

\noindent \textbf{Pruning.}
Pruning is originally invented as a method to reduce the size of large-scale neural networks\cite{han2015learning}. \cite{uchida2017embedding} utilized pruning as a method for removal of the watermark. Pruning eliminates a certain percentage of weighs having smaller absolute values from the neural network. After pruning, the entire network is retrained with a small number of training samples to recover the prediction accuracy.
\cite{uchida2017embedding,merrer2017adversarial,rouhani2018deepsigns,zhang2018protecting} employed pruning as a method to invalidate watermarking.

\vspace{0.15cm}

Since pruning with $0$ \% pruning rate corresponds to re-training, we consider pruning only in the following. 

\vspace{0.15cm}

\subsubsection{Performance evaluation of invalidation.}

The unauthorized service provider wishes to avoid verification of watermark by the model owner.
In this sense, AUC defined in Section \ref{sec:auc} can be a performance measure of invalidation (lower is better for the unauthorized service provider, higher is better for the model owner). 

At the same time, the unauthorized service provider does not wish that the prediction accuracy of the model is degraded by invalidation.
Thus, the test accuracy of $\model_\keySet$ under watermark invalidation can be used as the performance measure of invalidation (higher is better for the unauthorized model user). If the test accuracy of $\model_\keySet$ without invalidation is close to the test accuracy of the model with invalidation, it means that the invalidation does not degrade the prediction performance of the model, which is preferable to the unauthorized service provider.

\subsection{Problem Statement}

Watermarking of a neural network and its verification is processed by following the procedure below.

\begin{enumerate}
\item Embedding. A model owner embeds a watermark (key samples) $\keySet$ into her model $\model$ as $\model_\keySet$ by $\model_\keySet = \embed(\model, \keySet)$
\item Model distribution. An unauthorized service provider illegally obtains the model and provides prediction service online without permission
\item Verification: The model owner runs the verifiction process $\verify (\model, \keySet', \tau_{\mbox{acc}}))$ following three steps
  \begin{enumerate}
  \item Queries for verification: The model owner finds a suspicious prediction service and issues prediction queries: $\keySample_1,\keySample_2, \hdots, \keySample_{|\keySet'|}$ in $\keySet' \subseteq \keySet$ for watermark verification
  \item Invalidation of verification: Model user returns invalidated responses to the model owner as $\tilde{t}_1, \hdots, \tilde{t}_{|\keySet'|}$ where $\tilde{t}_{i} = \argmax_j \tilde{y}_{i,j}$ and $\tilde{\bi{y}}_i = \invalid(\model_\keySet (\keySample_i))$ for $i \in [|\keySet'|]$. Here, $\invalid$ can be either of model modification or query modification (introduced in the next section)
  \item Judgement: The model owner evaluates watermark accuracy ${\rm acc}_{f}^{\keySet'}$ using $\{\keyLabel_{i}\}_{i=1}^{|K'|}$ and $\{\tilde{t}_i\}_{i=1}^{|K'|}$ and judges whether or not ${\rm acc}_{g}^{\keySet'} > \tau_{acc}$ holds. If it holds, the model owner verifies the watermark.
  \end{enumerate}
\end{enumerate}

Embedding, verification, and invalidation of the watermark are thus defined as a game between the model owner and the unauthorized service provider. From the viewpoint of the model owner, she wants an embedding function with better predictive performance (test accuracy) and a verification function that achieves a higher AUC (watermark accuracy). On the other hand, from the viewpoint of the unauthorized service provider, he wants an invalidation function that does not significantly degrade the predictive performance (test accuracy) of the target model while he wants that the invalidation process degrades the AUC of verification as much as possible.

\section{Watermark invalidation by query modification}\label{sec:collapsed}

In this section, we introduce a novel watermark invalidation method, query modification, and demonstrate that existing watermarking methods can be significantly collapsed by either of model modification or query modification.

Our query modification is composed of two steps: key sample detection (Section \ref{sec:queryDetection}) and query modification by autoencoder (Section \ref{sec:queryModification}).
\begin{enumerate}
\item Key sample detection step. Given a query, the unauthorized service provider judges whether or not the query issued by someone works as a key sample for watermark verification.
\item Query modification step. If the query is judged as a key sample at the detection step, the key sample is modified so that it hinders the verification process. If not, no modification is made on the query.
\end{enumerate}

We suppose the unauthorized service provider can obtain a small number of samples drawn from the training sample distribution for query modification. We remark that existing model modification methods, such as pruning or re-training, assumes that a small number of samples are available as well.

Our query modification described in the following heavily relies on autoencoder. We first explain the functionality of autoencoder in Section \ref{sec:aec}.
Then, in Section \ref{sec:queryDetection}, we introduce two criteria to detect key samples using autoencoder and show that the method can detect various types of key samples generated by existing watermarking methods with high probability.
Next, in Section \ref{sec:eval}, we compare model modification and query modification regarding watermark accuracy (true positive rate).
Putting everything together, we obtain the full query modification algorithm, that detects key samples and collapses the verification process by deactivating the key samples selectively.

\subsection{Autoencoder}\label{sec:aec}

Autoencoder is a particular type of neural network which is commonly used to find a low-dimensional representation of high-dimensional samples in an unsupervised manner\cite{ae}.
Autoencoder is defined as a nonlinear map from the sample space to the sample space $\aec:\mathbb{R}^d \rightarrow \mathbb{R}^d$. Autoencoder first compresses high-dimensional samples into a low-dimensional latent space and then decompresses the low-dimensional signals into the sample space again so that the resulting samples closely matches the input.

When an autoencoder is trained with samples drawn from a distribution, outputs of the autoencoder are known to fit the latent representation of the distribution. Using this functionality of autoencoder, when key samples are created by superimposing specific images to original images, autoencoder can dilute the superimposed image to some extent. Also, when key samples are created by adding random noise, autoencoder can eliminate the noise to some extent. In this way, modification of key samples by autoencoder is expected to eliminate the effect of key samples and hinder the verification process of the model owner.

\begin{figure*}[t]
\centering
\subfigure[original]{\includegraphics[width=0.09\linewidth]{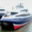}\label{fig:ordinal}}
\subfigure[\cite{zhang2018protecting}-content \newline ``test'', gray]{\includegraphics[width=0.09\linewidth]{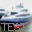}\label{fig:test(gray)}}
\subfigure[\cite{zhang2018protecting}-content \newline ``test'', blue]{\includegraphics[width=0.09\linewidth]{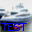}\label{fig:test(blue)}}
\subfigure[\cite{zhang2018protecting}-content \newline symbol]{\includegraphics[width=0.09\linewidth]{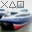}\label{fig:symbol}}
\subfigure[\cite{zhang2018protecting}-content \newline ``heart'', gray]{\includegraphics[width=0.09\linewidth]{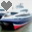}\label{fig:heart(gray)}}
\subfigure[\cite{zhang2018protecting}-content \newline ``heart'', red]{\includegraphics[width=0.09\linewidth]{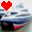}\label{fig:heart(red)}}
\subfigure[\cite{zhang2018protecting}-unrelated]{\includegraphics[width=0.09\linewidth]{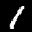}\label{fig:unrelated}}
\subfigure[\cite{zhang2018protecting}-noise]{\includegraphics[width=0.09\linewidth]{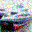}\label{fig:noise}}
\subfigure[\cite{merrer2017adversarial}-AFS]{\includegraphics[width=0.09\linewidth]{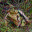}\label{fig:AFS}}
\subfigure[\cite{rouhani2018deepsigns}-DS]{\includegraphics[width=0.09\linewidth]{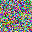}\label{fig:DS}}

\subfigure[$\aec$(a)]{\includegraphics[width=0.09\linewidth]{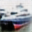}\label{fig:ordinal-ae}}
\subfigure[$\aec$(b)]{\includegraphics[width=0.09\linewidth]{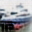}\label{fig:test(gray)-ae}}
\subfigure[$\aec$(c)]{\includegraphics[width=0.09\linewidth]{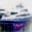}\label{fig:test(blue)-ae}}
\subfigure[$\aec$(d)]{\includegraphics[width=0.09\linewidth]{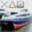}\label{fig:symbol-ae}}
\subfigure[$\aec$(e)]{\includegraphics[width=0.09\linewidth]{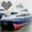}\label{fig:heart(gray)-ae}}
\subfigure[$\aec$(f)]{\includegraphics[width=0.09\linewidth]{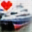}\label{fig:heart(red)-ae}}
\subfigure[$\aec$(g)]{\includegraphics[width=0.09\linewidth]{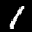}\label{fig:unrelated-ae}}
\subfigure[$\aec$(h)]{\includegraphics[width=0.09\linewidth]{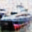}\label{fig:noise-ae}}
\subfigure[$\aec$(g)]{\includegraphics[width=0.09\linewidth]{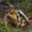}\label{fig:AFS-ae}}
\subfigure[$\aec$(h)]{\includegraphics[width=0.09\linewidth]{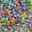}\label{fig:DS-ae}}

\caption{The examples of key samples of existing watermark method (top) and after applying autoencoder each top examples (bottom). The example (a) is an original image from CIFAR10 that used to making examples of (b)-(f) and (h) by \cite{zhang2018protecting}-content method and the examples of (g),(h),(i) and (j) are key samples of \cite{zhang2018protecting}-unrelated, \cite{zhang2018protecting}-noise, \cite{merrer2017adversarial}-AFS and \cite{rouhani2018deepsigns}-DS method respectively.}
\label{fig:aex}
\end{figure*}

We tested the effect of autoencoder with key samples generated with nine different methods.
We employed 6-layer convolutional autoencoder for this experiment. The detailed structure is shown in Table \ref{tb:aec-arch} in Apeendix \ref{ap:arch}.
\fig{aex} shows examples of key samples created from CIFAR10 dataset\cite{cifar10}.
\fig{ordinal} is an example of original images; \fig{aex}(b)-(j) are key samples generated with \fig{ordinal} by methods presented in \cite{zhang2018protecting}, \cite{merrer2017adversarial}, and \cite{rouhani2018deepsigns}. 

We trained an autoencoder with 5000 samples of CIFAR10 and applied the autoencoder to key samples. Key samples after application of autoencoder are shown in the second row of \fig{aex}. As we see from the figures, the superimposed images in \fig{aex}(b)-(f) are removed or diluted in \fig{aex}(l)-(p). Also, noise in \fig{aex}(h)-(i) is weakened in \fig{aex}(r)-(s).

\subsection{Key Sample Detection}\label{sec:queryDetection}

As we see in the last subsection, autoencoder can eliminate or dilute superimposed images for certain types of key samples. One straightforward attack against watermark is to apply autoencoder to every query. Unfortunately, this does not work well in reality because application of autoencoder to non-key samples degrades the prediction accuracy of the model significantly.
Considering that key samples issued by the model owner might be blended in a large number of regular queries issued by non-model owners at prediction time, application of autoencoder to every query would spoil the predictive performance of the prediction service of the unauthorized service provider.

More intelligent invalidation is to detect queries with key samples among queries with non-key samples and apply autoencoder only to the detected key samples. If this is possible, the degradation of predictive performance can be minimized. In this subsection, we show that key samples generated by existing methods can be detected with high probability.

\subsubsection{Key sample detection by reconstruction loss}

As we see in the last subsection, superimposed images or noise of key samples can be eliminated or diluted by autoencoder to some extent. This observation shows that this type of key samples can be detected by checking if autoencoder makes a significant change in the input image.

Let $\aec$ be an autoencoder trained with samples drawn from a training sample distribution of the target classifier. 
Then, the reconstruction loss of  $\sample$ introduced by the autoencoder is defined by
\begin{equation}
  \recLoss{\sample} = ||\sample-\aec(\sample)||_{2}^{2}
\end{equation}
where $\aec(\sample)$ is the sample reconstructed with the autoencoder and $\| \cdot \|_2$ denotes the $\ell_2$ norm.
If $\sample$ is an ordinary sample drawn from the training sample distribution, the reconstruction loss would be small. However, if $\sample$ is a key sample created by making a certain modification on an ordinal image, the reconstruction loss becomes relatively larger because the sample does not follow the training sample distribution. We can identify key samples by using this gap. More specifically, we judge $\sample$ as a key sample if $\recLoss{\sample}>\tau_{\rm REC}$ where $\tau_{\rm REC}$ is a threshold parameter determined in advance.

The threshold $\tau_{\rm REC}$ is determined empirically by the $r$ percentile point of the distribution of the reconstruction loss~\footnote{The cut point is a tunable parameter by the unauthorized service provider. If the unauthorized service provider wants to judge key samples more conservatively at the sacrifice of the test accuracy, the quantile point with a lower cut point can be used as the threshold. }. In the following experiments, we employed the $95\%$ quantile point as the threshold.

\subsubsection{Key sample detection by Jensen-Shanon divergence}

Detection by the reconstruction loss pays attention to the change in the sample space before and after the application of autoencoder. 
Next, we focus on the change of the predicted label distribution before and after application of autoencoder.
We measure the difference between $\model_K(\sample)$ and $\model_K(\aec(\sample))$.
Recall that key samples except \cite{merrer2017adversarial}-AFS are generated so that the label of the key samples does not represent the content of the images or does not agree with the true label of the original images for verification purpose.
Therefore, if the input is a key sample, the difference between $\model_K(\sample)$ and $\model_K(\aec(\sample))$ would become large. We use this difference to detect key samples.

Since $\model(\sample)$ forms a probability vector, we measure this difference by the Jensen-Shanon divergence (JSD):
\begin{equation}
  \begin{split} \jsd(\model_\keySet(\sample)||\model_\keySet(\aec(\sample))) = & \frac{1}{2}\kl(\model_\keySet(\sample)||B) \\
    & + \frac{1}{2}\kl(\model_\keySet(\aec(\sample))||B)\label{eq:jsd}
    \end{split}
\end{equation}
where $\kl$ is Kullback-Leibler divergence
$\kl(\model_\keySet(\sample)||\model_\keySet(\aec(\sample))) = \sum_{i} \model_\keySet(\sample)_{i} \log  \frac{ \model_\keySet(\aec(\sample))_{i}}{\model_\keySet(\sample)_{i}}$
and $B=(\model_\keySet(\sample)+\model_\keySet(\aec(\sample)))/2$.

The threshold for detection $\tau_{\mbox{JSD}}$ is also obtained as the empirical estimation of the $r$ percentile point of $\jsd(\model_\keySet(\sample)||\model_\keySet(\aec(\sample)))$ for the sample set given to the unauthorized service provider.

\subsubsection{Experimental evaluation of detection rate}

We experimentally evaluate the detection rate of key samples using the two criteria.

{\bf Experimental setup.} As the target classification model, we employed ResNet32~\cite{resnet}. We generated nine types of key samples listed in Section \ref{sec:embed} for MNIST\cite{mnist}, GTRSB\cite{gtsrb}, CIFAR10, and CIFAR100\cite{cifar10} datasets and embedded them into the model for watermark independently.
See Section \ref{sec:dataset} for the details of the dataset.

For watermark using \cite{zhang2018protecting}-content and \cite{zhang2018protecting}-noise, we labeled all key samples as ``0'' in MNIST, ``20 km/s speed-limit sign'' in GTSRB, ``airplane'' in CIFAR10, and ``apple'' in CIFAR100.
For \cite{zhang2018protecting}-unrelated, we used images of ``m'' (2,400 samples) in EMNIST \cite{emnist} labeled with ``0'' as key samples for MNIST. Also, we used images of ``1'' in MNIST (6,742 samples) as key samples with label ``20 km/s speed-limit sign'', ``airplane'', and ``apple'', for GTSRB, CIFAR10 and CIFAR100, respectively.

Since embedding by \cite{zhang2018protecting} requires that the model well generalizes the symbols or logotypes, it requires a large number of key samples to embed.
Watermarking using \cite{merrer2017adversarial}-AFS and \cite{rouhani2018deepsigns}-DS take advantage of model overfitting to specific key samples, we used a relatively smaller number of key samples (30 key samples) for these methods. We experimentally confirm that if we use a larger number of key samples for these methods, it degrades the test accuracy due to overfitting of the model.
For more detail about sample assignment, see \tb{num_samples} in Appendix \ref{ap:assignment}.

The threshold $\tau_{\rm REC}$ and $\tau_{\rm JSD}$ were determined as the 95 percentile point of the reconstruction loss and the Jensen-Shannon divergence. 10-fold cross-validation with the samples possessed by the unauthorized user was used for estimation.

{\bf Results.} Given key samples generated by each method, we evaluated the detection rate in Table \ref{tb:detectability}.
In the table, the first low ``rec loss'' shows the rate that key samples are correctly judged as key samples when $\recLoss{\sample}>\tau_{\rm REC}$ is used as the detection rule.
The second low ``JSD'' shows the detection rate when $\jsd(\model_\keySet(\sample)||\model_\keySet(\aec(\sample)))>\tau_{\rm JSD}$ is used as the detection rule.
The third low ``both'' shows the detection rate when samples satisfying either of $\recLoss{\sample}>\tau_{\rm REC}$ or $\jsd(\model_\keySet(\sample)||\model_\keySet(\aec(\sample)))>\tau_{\rm JSD}$ are classified as key samples.
Since MNIST dataset consists of gray scale images only, results for \cite{zhang2018protecting}-content(test, blue) and \cite{zhang2018protecting}-content(heart, red) are not shown.
As we see from the results, most of the key samples for CIFAR10 and CIFAR100 can be detected with high probability when we use both the reconstruction loss and JSD.
Also, not all but some of the key samples for MNIST and GTSRB can be detected with high probability, too.

\begin{table*}[t]
  \begin{center}
    \caption{Detection rate of key samples. The lows ``rec loss'' and ``JSD'' show the detection rate when reconstruction loss and JSD are used for the detection rule, respectively. ``both'' shows the detection rate when both measures are used for detection.}
    \label{tb:detectability}
    \scalebox{0.8}{
      \begin{tabular}{c|ccccccccc} \hline
        \multicolumn{10}{c}{MNIST} \\ \hline
        & \multicolumn{5}{c}{\cite{zhang2018protecting}-content} & & & & \\
        mesuears$\backslash$key samples & ``test'', gray & ``test'', blue & symbol & ``heart'', gray & ``heart'', red & \cite{zhang2018protecting}-unrelated & \cite{zhang2018protecting}-noise & \cite{merrer2017adversarial}-AFS & \cite{rouhani2018deepsigns}-DS \\ \hline
        rec loss & 1.0 & - & 1.0 & 1.0 & - & 0.28 & 0.49  & 0.0  & 1.0 \\
        JSD      & 0.0 & - & 1.0 & 1.0 & - & 0.0  & 0.067 & 0.23 & 1.0 \\
        both     & 1.0 & - & 1.0 & 1.0 & - & 0.28 & 0.49  & 0.23 & 1.0 \\ \hline
        \multicolumn{10}{c}{GTSRB} \\ \hline
        & \multicolumn{5}{c}{\cite{zhang2018protecting}-content} & & & & \\
        mesuears$\backslash$key samples & ``test'', gray & ``test'', blue & symbol & ``heart'', gray & ``heart'', red & \cite{zhang2018protecting}-unrelated & \cite{zhang2018protecting}-noise & \cite{merrer2017adversarial}-AFS & \cite{rouhani2018deepsigns}-DS \\ \hline
        rec loss & 0.10  & 0.95 & 0.063 & 0.047 & 0.09 & 0.12 & 0.51  & 0.0   & 1.0 \\
        JSD      & 0.067 & 0.04 & 0.32  & 0.18  & 0.01 & 0.0  & 0.0   & 0.033 & 0.26 \\
        both     & .14   & 0.95 & 0.37  & 0.21  & 0.1  & 0.12 & 0.51  & 0.033 & 1.0 \\ \hline
        \multicolumn{10}{c}{CIFAR10} \\ \hline
        & \multicolumn{5}{c}{\cite{zhang2018protecting}-content} & & & & \\
        mesuears$\backslash$key samples & ``test'', gray & ``test'', blue & symbol & ``heart'', gray & ``heart'', red & \cite{zhang2018protecting}-unrelated & \cite{zhang2018protecting}-noise & \cite{merrer2017adversarial}-AFS & \cite{rouhani2018deepsigns}-DS \\ \hline
        rec loss & 0.55 & 1.0 & 0.52 & 0.09 & 0.98 & 0.27 & 1.0 & 1.0 & 1.0 \\
        JSD      & 0.95 & 0.95 & 0.96 & 0.97 & 0.18 & 0.0 & 0.65 & 0.2 & 0.66 \\
        both     & 0.98 & 1.0 & 0.98 & 0.99 & 0.98 & 0.27 & 1.0 & 1.0 & 1.0 \\ \hline
        \multicolumn{10}{c}{CIFAR100} \\ \hline
        & \multicolumn{5}{c}{\cite{zhang2018protecting}-content} & & & & \\
        mesuears$\backslash$key samples & ``test'', gray & ``test'', blue & symbol & ``heart'', gray & ``heart'', red & \cite{zhang2018protecting}-unrelated & \cite{zhang2018protecting}-noise & \cite{merrer2017adversarial}-AFS & \cite{rouhani2018deepsigns}-DS \\ \hline
        rec loss & 0.49 & 1.0 & 0.53 & 0.12 & 0.99 & 0.0 & 1.0 & 0.43 & 1.0 \\
        JSD      & 0.81 & 0.71 & 0.79 & 0.94 & 0.05 & 0.03 & 0.26 & 0.1 & 0.1 \\
        both     & 0.92 & 1.0 & 0.91 & 0.96 & 0.99 & 0.037 & 1.0 & 0.47 & 1.0 \\ \hline
      \end{tabular}
      }
  \end{center}
\end{table*}

\subsection{Query modification}\label{sec:queryModification}

In this subsection, we compare watermark invalidation by query modification and model modification.

\subsubsection{Evalaution of watermark accuracy after query modification and model modification}\label{sec:eval}

We compare the watermark accuracy when the unauthorized service provider returns
\begin{itemize}
\item $\model_\keySet(\keySample)$ without model and query modification,
\item $\model_\keySet(\aec(\keySample))$ with query modificaion by autoencoder, and
\item $\tilde{\model}_\keySet(\keySample)$ with model modificaion by pruning.
\end{itemize}

{\bf Experimental setup.}
In query modification, given a query $\keySample$, the unauthorized user returns $\model_\keySet(\aec(\keySample))$. For this evaluation, we did not use the detection step because all given samples are key samples to evaluate the true positive rate.

In model modification, we first prune $q$ percentage of weights having smaller absolute values in the model is made zero (pruning). We changed $q$ from $0$ to $90$ by $10$.
After pruning, the entire network is retrained for ten epochs by Adam with learning rate $0.001$.
When the pruning rate $q$ becomes larger, model verification can become more difficult while the test accuracy of the model is decreased more. Since it is meaningless to attain watermark invalidation with sacrificing test accuracy, we varied pruning rate $r$ so that the deterioration of the test accuracy is less than 10\% of the baseline (e.g, test accuracy of the model without watermark). For each setting, we tuned the pruning rate so that the watermark accuracy becomes the worst. 

\begin{table*}[t]
  \begin{center}
    \caption{Watermark accuracy before (no invalidation) and after applying queri modification (autoencoder) and model modification (pruning). }
    \label{tb:ae_key_acc}
    \scalebox{0.8}{
      \begin{tabular}{cc|ccccccccc} \hline
        
        & & \multicolumn{5}{c}{\cite{zhang2018protecting}-content} & & & & \\
    dataset    & watermark invalidation  $\backslash$key samples & ``test'', gray & ``test'', blue & symbol & ``heart'', gray & ``heart'', red & \cite{zhang2018protecting}-unrelated & \cite{zhang2018protecting}-noise & \cite{merrer2017adversarial}-AFS & \cite{rouhani2018deepsigns}-DS \\ \hline
              & no invalidation & 1.0   & -   & 1.0 & 1.0 & - & 1.0 & 1.0 & 1.0 & 1.0 \\
        MNIST & query mod. (autoencoder) & 0.99   & -   & 0.0007 & 0.96 & - & 1.0 & 1.0 & 1.0 & 0.4 \\
              & model mod. (pruning) & 0.0034 & - & 0.023 & 0.0020 & - & 0.89 & 0.20 & 1.0 & 0.67 \\ \hline
              & no invalidation & 1.0   & 1.0   & 1.0 & 1.0 & 1.0 & 1.0 & 1.0 & 1.0 & 1.0 \\
        GTSRB & query mod. (autoencoder) & 0.39 & 0.20 & 0.019 & 0.081 & 0.28 & 0.78 & 0.051 & 0.90 & 0.10 \\
              & model mod.  (pruning)  & 0.025 & 0.0011 & 0.0012 & 0.0 & 0.0011 & 0.0 & 0.066 & 0.87 & 0.1 \\ \hline 
              & no  invalidation & 1.0   & 1.0   & 1.0 & 1.0 & 1.0 & 1.0 & 1.0 & 1.0 & 1.0 \\        
        CIFAR10 & query mod. (autoencoder) & 0.069 & 0.021 & 0.017 & 0.021 & 0.39 & 1.0   & 0.014  & 0.73 & 0.23 \\ 
                & model mod.  (pruning)  & 0.57  & 0.73  & 0.30  & 0.99  & 0.14 & 0.010 & 0.070  & 0.86 & 0.30 \\ \hline 
              & no invalidation & 1.0   & 1.0   & 1.0 & 1.0 & 1.0 & 1.0 & 1.0 & 1.0 & 1.0 \\        
        CIFAR100 & query mod. (autoencoder) & 0.064 & 0.026 & 0.010 & 0.016 & 0.35 & 0.68 & 0.0082 & 0.56 & 0.0 \\
        & model mod. (pruning)  & 0.88 & 0.99 & 0.082 & 0.95 & 0.96 & 0.11 & 0.16 & 0.57 & 0.17 \\ \hline 
      \end{tabular}
      }
  \end{center}
\end{table*}

{\bf Results.}
The results are summarized in Table \ref{tb:ae_key_acc}.
The table compares the watermark accuracy when nine types of key samples are used watermark verification under query modification (autoencoder) and model modification (pruning) for four datasets.
Every type of key samples achieves watermark accuracy 1.0 for all datasets when watermark invalidation by model or query modification is not applied.
This means that watermark works perfectly if the unauthorized service provider does not try to invalidate the watermark of the model.
The watermark accuracy under query modification significantly decreases for CIFAR10B and CIFAR100 with most of the key samples.
Considering that the detection of key samples works more successfully for CIFAR-type tasks, query modification works as a strong attack against CIFAR-type tasks.
On the other hand, for MNIST and GTSRB, attack by model modification works successfully.
Interestingly, model modification and query modification work complementary to each other.
The results show that there exist no key samples that can achieve high watermark accuracy under both model modification and query modification for all datasets.
In Section \ref{sec:proposal}, we introduce a novel watermarking method that can achieve high watermark accuracy under both model modification and query modification.

\subsection{Invalidation by query modification}\label{sec:invalidationByQM}

We confirmed that, for certain types of key samples, our detection method could identify them autoencoder can invalidate the watermark.
Putting everythin together, we summarize the entire invalidation process using autoencoder, which is described by the following procedure:

\vspace{0.15cm}
{\bf Query modification by autoencoder}
\begin{enumerate}
\item Require: Training sample set $\sampleSet$, parameter $r$ (percentile), $k$ (cross validation)
\item Preparation:
  \begin{enumerate}
  \item Train autoencoder $\aec$ with $\sampleSet$
  \item Evaluate the $r$-percentile quantile point of $\recLoss{\sample}$ where $\sample \in \sampleSet$ by $k$-fold cross validation and set the $r$-percentile quantile point as $\tau_{\rm REC}$
  \item Evaluate the $r$-percentile quantile point of $\jsd(\model_\keySet(\sample)||\model_\keySet(\aec(\sample)))$ where $\sample \in \sampleSet$ by $k$-fold cross validation and set the $r$-percentile quantile point as $\tau_{\rm JSD}$
  \end{enumerate}
\item Key sample detection: Given a query $\sample$, if $\jsd(y||\hat{y}) > \tau_{\rm JSD}$ or $\jsd(y||\hat{y}) > \tau_{\rm JSD}$, judge $\sample$ as a key sample. Otherwise, judge $\sample$ as an ortinal sample.
\item Query modification: If $\sample$ is detected as a key sample, return $\model(\aec(\sample))$ to the model owner. Otherwise, return $\model(\sample)$ to the model owner.
\end{enumerate}    
\vspace{0.15cm}

In the preparation step, the unauthorized user trains an autoencoder and estimate the two thresholds using samples. Upon request of prediction with a query, the unauthorized user judges if the query works as a key sample at detection step. If so, the query is modified by the autoencoder so that it does not work as a key sample and returns $\model(\aec(\sample))$ as the prediction result. Otherwise, it returns $\model(\sample)$ without any modification.

\section{Wartermarking with Exponential Weighting}\label{sec:proposal}

In this section, we propose a watermarking method that resists invalidation with both model modification and query modification more robustly.
Our defense against watermark invalidation consists of two strategies. One is a defense against query modification.
The other is a strategy for embedding the watermark into models that are tolerant of watermark invalidation.

\subsection{Generation of Key Samples by Label Change}

As we observed in the last section, key samples generated by existing methods can be invalidated by either of model modification or query modification.
As a defense against query modification, we need a key sample that cannot be detected by unauthorized users while it can be verified by the model owner.

Query modification detects a query that largely deviates from the training sample distribution as a key sample and then modifies the detected key sample so that it follows the training sample distribution by autoencoder. Our idea to avoid this invalidation is to utilize a sample that perfectly follows the training sample distribution as a key sample.
Since what the unauthorized user observes is non-labeled queries only, such key samples following the training sample distribution are undetectable. We embed such key samples into the model as the watermark.

In the generation process of a key sample, we randomly select a training sample without any modification, and we then change the label of the sample to a label that is different from original one (label change).
For example, we label an image of a dog as ``cat'' and use this as a key sample.
If we train a model with a training dataset containing key samples generated in this way, the model would recognize dog images except for this specific key sample as ``dog'' while this key sample of the dog image is recognized as ``cat''. The model owner can verify watermark using this response from the model.
Recall that queries issued by model owners are not labeled. Thus, key samples generated in such a way become perfectly indistinguishable from training samples. Once such key samples are embedded into a model, the model owner can verify the watermark without being detected by the unauthorized service provider.

Our key samples are somewhat similar to \cite{zhang2018protecting}-unrelated.  \cite{zhang2018protecting}-unrelated selects images from unrelated classification task. These can be detected by the unauthorized service providers who train the generative model of the training samples, as we did by autoencoder.

\subsection{Embedding Key Samples with Exponential Weighting}

Key samples generated by label change are undetectable. However, training with such key samples would cause overfit to the model.
Our preliminary experiments revealed that watermark embedded in this way could be instantly invalidated by model modification, such as pruning or re-training.
This is because pruning or re-training resolves to overfit of the model to key samples. 

If a large number of model parameters having small absolute values are involved in prediction, prediction results of samples would be significantly changed by pruning or re-training of the model.
Our idea to avoid this by imprinting key samples with greater force so that they are not removed by model modification. More specifically, during the training process, we identify model parameters of the neural network model that significantly contribute to giving predictions and increase the weight value exponentially so that model modification cannot change the prediction behavior of samples (including key samples) before and after model modification.

In the $l$th layer of model $f$, we denote the input to the $l$th layer by $h^{l}$ and the model parameters by $\theta^{l}$. $op^{l}(h^{l},\theta^{l})$ denotes the operetor to process computation with $h^{l}$ and $\theta^{l}$ in the $l$th th layer (e.g. Affine transformation, convolution). $a^{l}$ denotes the activation function.
Then, the output of the $l$th layer is given by
\begin{equation}\label{eq:op}
 h^{l+1} = a^{l}(op^{l}(h^{l},\theta^{l})).
\end{equation}
where the output is transfered to the input of the $l+1$th layer, $h^{l+1}$.

Let us denote the $i$th element of $\theta^{l}$ as $\theta^{l}_{i}$.
We exponentially weight each model parameter as follows.
\begin{equation}\label{eq:ew}
 EW(\theta^{l},T) = \theta^{l}_{\rm exp}, \theta^{l}_{\rm exp,i} = \frac{\exp{|\theta^{l}_{i}|T}}{{\max_{i}{\exp{|\theta^{l}_{i}|T}}}} \theta^{l}_{i}
\end{equation}
where $T$ is a hyperparameter for adjusting the intensity of the weighting.
After weighting the parameters, the $l$th layer finally outputs
\begin{equation}\label{eq:opEW}
 h^{l+1} = a^{l}(op^{l}(h^{l},EW(\theta^{l},T))).
\end{equation}

The procedure of watermarking by exponential weighting is as follows.

\vspace{0.15cm}
{\bf Watermarking by exponetial weighting}
\begin{enumerate}
\item Require: Training sample set $X$, key sample set $K$ generated with label change
\item Train the model $f$ with $X$ where we denote operation of each layer by \eq{op}
\item After training, replace \eq{op} of each layer of the model $f$ with \eq{opEW}. Then, retrain $f$ with $X \cup K$ and obtain $f_K$
\end{enumerate}
\vspace{0.15cm}

We explain forward and back-propagation of neural networks represented by \eq{opEW} with respect to $\theta^{l}$.
In forward-propagation, $op^{l}$ is performed with model parameters exponentially weighted by $EW$. With this, parameters that do not have large absolute values are forced to have small values. Since parameters with small values do not have a large influence on operation $op^{l}$, only parameters that have large absolute values influence operation $op^{l}$ eventually.
In back-propagation, gradients of $h^{l+1}$ with respect to $\theta^{l}$ is calculated using the differential chain rule as follows.
\begin{equation}
  \frac{\partial h^{l+1}}{\partial \theta^{l}} = \frac{\partial h^{l+1}}{\partial op^{l}} \frac{\partial op^{l}}{\partial EW(\theta^{l})} \frac{\partial EW(\theta^{l})}{\partial \theta^{l}}\label{eq:ew}
\end{equation}
Exponential weighting can be treated as one of activation functions of neural netwokrs. Back-propagation with exponential weighting is incorporated into back propagation in a natural manner as in Eq. \ref{eq:ew}.

\section{Experiments}\label{sec:exp}

In this section, we evaluate the proposed watermarking method regarding the predictive performance of the model after embedding watermarks (test accuracy) and the verification performance (AUC) under model modification and query modification.
The results are compared with the existing method~\cite{zhang2018protecting, merrer2017adversarial, rouhani2018deepsigns}.
Comparison with \cite{adi2018turning} is not shown because this method assumes the existence of a trusted third party and is not directly comparable.

\subsection{Experimental setting}

\subsubsection{Dataset}\label{sec:dataset}
We use the following four image datasets (MNIST, GTSRB, CIFAR10, and CIFAR100) for the evaluation.
Pixel values of images in all datasets are in $[0,1]$.

\textbf{MNIST}\cite{mnist} is a grayscale handwritten digits dataset that has a training set of 60,000 samples, and a test set of 10,000 samples.
The size of each image is $28 \times 28$ pixels, and there are ten classes from "0" to "9".

\textbf{GTSRB}\cite{gtsrb} is a RGB traffic sign dataset that consists of 39,209 training samples and 12,630 test samples. Each image is labeled with one of 43 classes.
Since the size of images contained in this dataset is not unique, we resize all images as $32 \times 32$ pixels.

\textbf{CIFAR10} and \textbf{CIFAR100} are RGB object classification datasets with 50,000 training samples and 10,000 test samples.
The size of each image is $32 \times 32$ pixels and is labeled with one of 10 and 100 classes, respectively.

Existing studies suppose the unauthorized service provider can use the training samples held by the model owner for watermark invalidation, which is not often possible in reality. In our experimental setup, we suppose training samples and samples used for watermark invalidation are mutually exclusive.
For the detailed setting about the assignment of samples for the model owner and the unauthorized service provider, see \tb{num_samples} in Appendix \ref{ap:assignment}.

\subsubsection{Model training and key sample generation and embedding}
We employed ResNet~32\cite{resnet} for the target classification model.
Also, we employed a six-layer convolutional autoencoder for query modification.
The detailed architecture of the autoencoder is summarized in Table \ref{tb:aec-arch} in Appendix \ref{ap:arch}.
\tb{param} in Appendix \ref{ap:arch} summarizes the hyperparameters used to train ResNet and autoencoder.

Key sample generation by existing methods follows Section \ref{sec:embed}.
We did not create key samples \cite{zhang2018protecting}-content ``test'', blue and \cite{zhang2018protecting}-content ``heart'', red in MNIST because images of MNIST dataset are gray scale.

For \cite{zhang2018protecting}-content, unrelated and noise methods, we embedded the watermark into the model by training the model with training samples and key samples from scratch.
For \cite{merrer2017adversarial}-AFS and \cite{rouhani2018deepsigns}-DS methods, we trained the model with only training samples and retrained with training samples and key samples to embed watermark the model.

For our watermarking method, we randomly selected 30 samples from the training dataset and generate key samples. For each key sample, we assign a label that is different from the original label randomly and embed them as watermark with exponential weighting.
Recall that watermarking of a neural network and its verification is a game between the model owner and the unauthorized service provider.
Parameters for invalidation methods (e.g., pruning rate, thresholds for key sample detection) are tuned so that the watermark accuracy becomes the lowest while parameters for embedding methods (e.g., the temperature parameter of exponential weighting) are tuned so that the watermark accuracy becomes the highest.
For exponential weighting, we adopted $T = 2.0$ for all datasets by preliminary experiments.
For evaluation of AUC, we used a model trained with training samples only.

\subsection{Predictive Performance Before and After Embedding Watermark}

We first evaluate the prediction performance of the model before and after embedding a watermark by test accuracy.
In \tb{inv-acc}, the columns with ``no inv.'' show the test accuracy of models with watermark when the unauthorized service provider performs no watermark invalidation.
As seen from the results of the columns ``no inv.'' in \tb{inv-acc}, all embedding methods can well preserve the predictive performance of the model even after embedding the watermark.

The columns with ``model mod.'' and ``query mod.'' in \tb{inv-acc} show the test accuracy of models with watermark under model modification (pruning) and query modification (autoencoder), respectively.
The difference of the test accuracies before and after watermark invalidation is at most 10\% in all settings.
Therefore, these invalidations are reasonable (i.e., does not degrade the test accuracy too much) from the viewpoint of the unauthorized service provider in our experiment.

\begin{table*}[t]
  \begin{center}
    \caption{The test accuracy of models without watermark and models watermarked by existing and the proposed method under watermark invalidation in four datasets. ``No inv'' denotes that no watermark invalidation is performed. ``model mod.'' and ``query mod.'' denotes model modification by pruning and query modification is performed for watermark invalidation. The number in the parenthesis in the ``model mod.'' column represents the pruning rate $r$ that gives the worst watermark accuracy.}
    \label{tb:inv-acc}
    \scalebox{0.8}{
      \begin{tabular}{|c|c|c|c|c|c|c|c|c|c|c|c|c|} \hline
        &\multicolumn{3}{c|}{MNIST} & \multicolumn{3}{c|}{GTSRB} & \multicolumn{3}{c|}{CIFAR10} & \multicolumn{3}{c|}{CIFAR100} \\ \cline{2-13}
        & \multicolumn{3}{c|}{invalidation} & \multicolumn{3}{c|}{invalidation} & \multicolumn{3}{c|}{invalidation} & \multicolumn{3}{c|}{invalidation}\\ \cline{2-13}
        methods & no inv.   & model mod.  & query mod. & no inv. & model mod.  & query mod. & no inv. & model mod.  & query mod. & no inv. & model mod.  & query mod. \\ \hline
        no watermark & 99.6\% & - & - & 97.3\% & - & - & 91.3\% & - & - & 66.9\% & - & - \\ \hline
        \cite{zhang2018protecting}-content   & 99.6\% & 99.4\% ($20\%$) & 94.0\% & 97.0\% & 97.4\% ($10\%$) & 93.0\% & 91.5\% & 89.9\% ($10\%$) & 89.1\%  & 67.6\% & 62.7\% ($0\%$) & 63.8\% \\ \hline
        \cite{zhang2018protecting}-content   & - & - & - & 97.0\% & 97.2\% ($0\%$) & 92.1\% & 91.0\% & 89.0\% ($40\%$) & 88.7\%  & 67.2\% & 62.3\% ($10\%$) & 62.9\% \\ \hline
        \cite{zhang2018protecting}-content   & 99.7\% & 99.3\% ($30\%$) & 99.4\% & 97.2\% & 97.6\% ($10\%$) & 93.7\% & 91.1\% & 90.6\% ($0\%$) & 88.8\%  & 67.8\% & 62.1\% ($10\%$) & 63.4\% \\ \hline
        \cite{zhang2018protecting}-content   & 99.7\% & 99.3\% ($0\%$) & 99.5\% & 97.2\% & 97.5\% ($10\%$) & 93.7\% & 91.4\% & 90.1\% ($10\%$) & 88.6\%  & 66.7\% & 61.9\% ($10\%$) & 62.3\% \\ \hline
        \cite{zhang2018protecting}-content   & - & - & - & 97.3\% & 97.8\% ($10\%$) & 93.7\% & 91.0\% & 88.1\% ($20\%$) & 89.0\%  & 67.9\% & 62.5\% ($0\%$) & 62.8\% \\ \hline
        \cite{zhang2018protecting}-unrelated & 99.6\% & 99.1\% ($20\%$) & 93.8\% & 97.0\% & 92.7\% ($90\%$) & 94.1\% & 90.7\% & 89.4\% ($0\%$) & 88.7\%  & 67.8\% & 57.1\% ($70\%$) & 65.1\% \\ \hline
        \cite{zhang2018protecting}-noise     & 99.6\% & 99.2\% ($20\%$) & 97.9\% & 97.0\% & 97.2\% ($0\%$) & 93.3\% & 91.5\% & 82.1\% ($80\%$) & 89.3\%  & 66.7\% & 55.3\% ($70\%$) & 64.2\% \\ \hline
        \cite{merrer2017adversarial}-AFS     & 99.6\% & 99.2\% ($10\%$) & 99.4\% & 97.7\% & 94.8\% ($90\%$) & 94.7\% & 92.0\% & 87.1\% ($80\%$) & 89.7\%  & 68.1\% & 57.4\% ($70\%$) & 65.9\% \\ \hline
        \cite{rouhani2018deepsigns}-DS       & 99.6\% & 99.1\% ($90\%$) & 99.4\% & 97.3\% & 95.4\% ($90\%$) & 94.7\% & 91.2\% & 86.6\% ($80\%$) & 89.3\%  & 68.3\% & 56.7\% ($70\%$) & 63.3\% \\ \hline
        proposal                             & 99.2\% & 99.0\% ($0\%$) & 91.9\% & 96.8\% & 95.0\% ($90\%$) & 93.2\% & 92.0\% & 82.5\% ($90\%$) & 90.3\%  & 67.4\% & 56.5\% ($80\%$) & 62.8\% \\ \hline
      \end{tabular}
      }
  \end{center}
\end{table*}

\subsection{Verification Performance under Invalidation}

We evaluate the verification performance of the watermarking methods by AUC under model modification and query modification.
Let $\model$ and $\model_K$ be model without and with watermark, respectively.
AUC is evaluated as follows:
\begin{enumerate}
\item Sample $\keySet'$ from $\keySet$ randomly with replacement.
\item For each $\keySample \in \keySet'$, evluate $\model(\keySample)$ and $\model_K(\keySample)$ as queries and evaluate the true positive rate and false positive rate 
\item We repeat Step 1 and Step 2 for 30 times, draw the ROC curve, and evaluate the AUC.
\end{enumerate}

As the baseline of the verification performance, we first evaluated the AUCs of the model without any watermark invalidation.
We experimentally confirmed that the AUCs of the verification performance with all the watermark methods except \cite{merrer2017adversarial}-AFS was $1$.
This means that the watermark can be perfectly verified by the model owner when the watermark is correctly embedded, and no watermark invalidation is performed.
The AUC of AFS was not one but still high. This is because models without watermark can predict correct labels for key samples (adversarial examples in AFS) sometimes and this makes the false positive rate non zero.

Next, we evaluated the verification performance of the model with watermark under watermark invalidation.

\begin{figure*}[t]
\centering
\def\subfigcapskip{-7pt}
\subfigure[AUC under mdel modification (pruning, top) and query modification (autoencoder, bottom) in $|K'|=20$]{\includegraphics[width=1.0\linewidth]{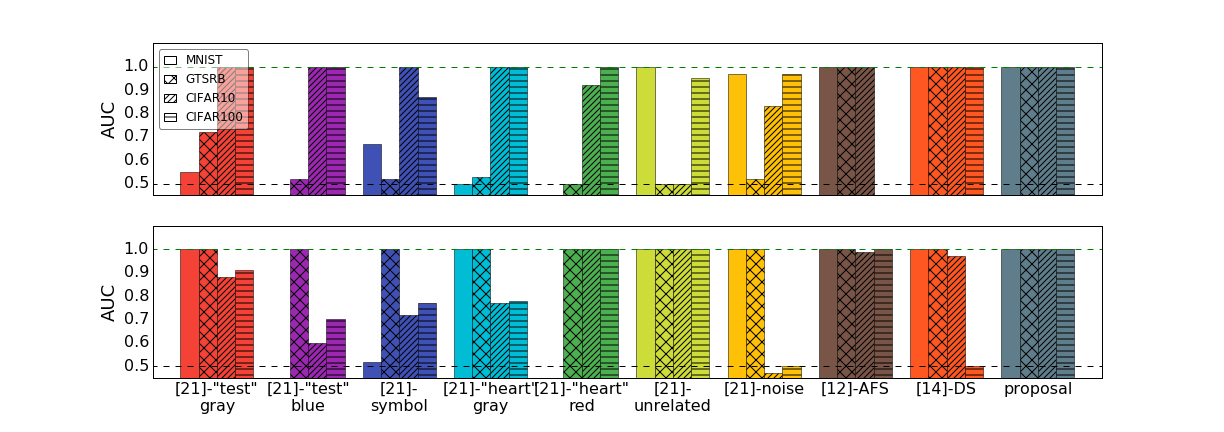} \label{fig:auc_K20}}
\subfigure[AUC under mdel modification (pruning, top) and query modification (autoencoder, bottom) in $|K'|=10$]{\includegraphics[width=1.0\linewidth]{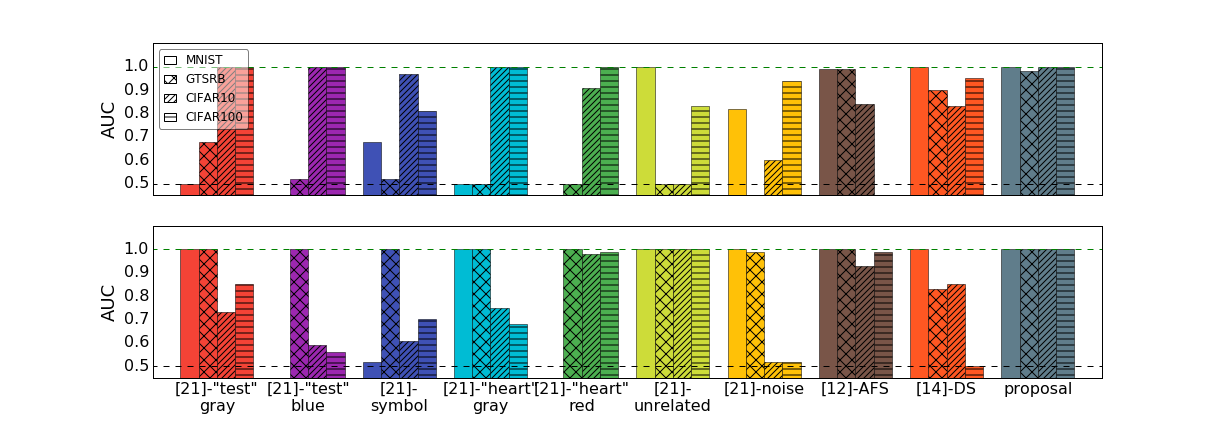} \label{fig:auc_K10}}
\subfigure[AUC under mdel modification (pruning, top) and query modification (autoencoder, bottom) in $|K'|=5$]{\includegraphics[width=1.0\linewidth]{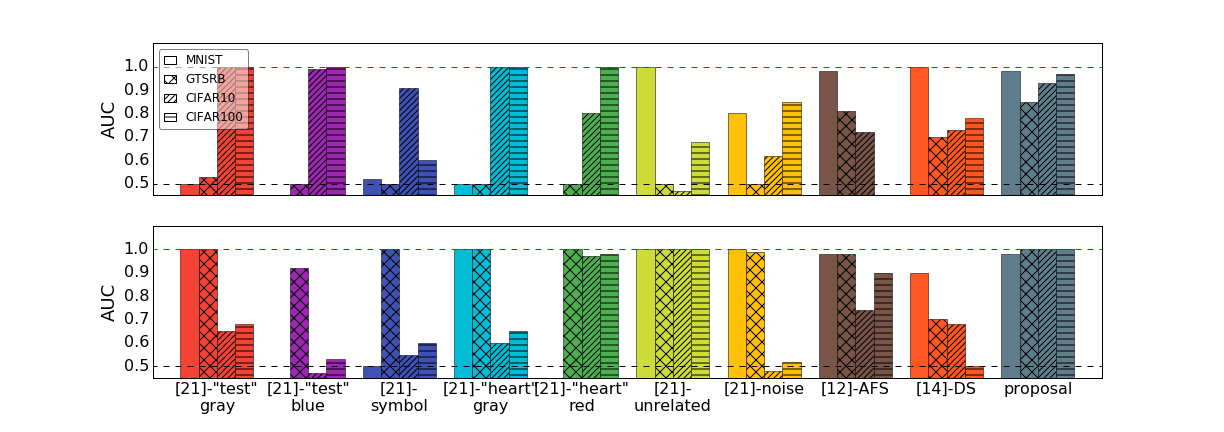} \label{fig:auc_K5}}
\caption{Verification performance (AUC) of existing and proposed watermark methods under invalidation by model modification (pruning, top of each subfigure) and query modification (bottom of each subfigure) for CIFAR10, CIFAR100, GTSRB and MNIST when $|K'|=20,10,5$. The vertical axis shows AUC and the horizontal axis shows the watermarking methods in each dataset. Black dash line shows ``AUC 0.5'' (verification is perfomed randomly) and green dash line shows ``AUC 1'' (verification performed perfectly).}\label{fig:auc}
\end{figure*}

\fig{auc} shows the AUCs of verification under invalidation by model modification and query modification where the number of key samples $|K'|$ is varied as $20,10,$ and $5$.

\fig{auc_K20} shows the AUC of each watermark method under model modification (pruning, top) and query modification (bottom) when $|K'|=20$.
In \cite{zhang2018protecting}-content type method, the tolerance of pruning depends on the datasets; AUC is high in CIFAR10 and CIFAR100 while AUC is low in GTSRB and MNIST.
The tolerance to query modification depends on the special image to be superimposed and the datasets. \cite{zhang2018protecting}-content ``test'' gray, and \cite{zhang2018protecting}-content``heart'' red achieve high AUC in all datasets even under query modification while the other types of key samples have low AUC in CIFAR10 and CIFAR100.
\cite{zhang2018protecting}-unrelated has high AUC in all datasets under query modification, but AUC is very low in CIFAR10 and GTSRB under invalidation by pruning.
\cite{zhang2018protecting}-noise has low AUC under pruning in GTSRB and query modification in CIFAR10 and CIFAR100.
\cite{merrer2017adversarial}-AFS has very low AUC under pruning in CIFAR100.
\cite{rouhani2018deepsigns}-DS has very low AUC under query modification in CIFAR 100.
The proposed method achieves AUC $1$ under any invalidation for all datasets.
Thus, no existing methods achieve AUC $1$ in all cases while the proposed method achieves AUC $1$ in all cases.
From these results, the proposed method achieves the best watermark accuracy when $|K'|=20$.

Next, we see the results at $|K'|=10$ and $|K'|=5$ (\fig{auc_K10} and \fig{auc_K5}).
The verification performance at $|K'|=10, 5$ is expected to be less than $|K'|=20$ because verification performance decreases when $|K'|$ is small.
In fact, for example, in the results of \cite{rouhani2018deepsigns}-DS, it achieves AUC $=1$ in seven of eight cases when $|K'|=20$ but AUC decreases as $|K'|$ decreases.
The proposed method achieves AUC $1$ in almost all cases even when $|K'|=10$. When $|K'|=5$, the worst AUC among all the settings is about 0.85 (under pruning in GTSRB), but it still achieves a higher AUC than AUCs of all existing methods.

As seen from the results, we can conclude that the proposed method, key samples with label change and embedding with exponential weighting, shows the best tolerance of both model modification and query modification without significant decrease of test accuracy.

\section{Conclusion}\label{sec:conc}

In this study, we proposed a novel watermarking method for the neural network that is tolerant of both model modification and query modification.
Our watermarking method consists of two components, (1) key generation by label change, and (2) key embedding by exponential weighting.

We experimentally demonstrated that our watermarking method achieves high verification performance even under a malicious attempt of unauthorized service providers to invalidate the verification process such as model modification and query modification.

\bibliographystyle{natbib}
\bibliography{reference}


\begin{thebibliography}{21}


\ifx \showCODEN    \undefined \def \showCODEN     #1{\unskip}     \fi
\ifx \showDOI      \undefined \def \showDOI       #1{#1}\fi
\ifx \showISBNx    \undefined \def \showISBNx     #1{\unskip}     \fi
\ifx \showISBNxiii \undefined \def \showISBNxiii  #1{\unskip}     \fi
\ifx \showISSN     \undefined \def \showISSN      #1{\unskip}     \fi
\ifx \showLCCN     \undefined \def \showLCCN      #1{\unskip}     \fi
\ifx \shownote     \undefined \def \shownote      #1{#1}          \fi
\ifx \showarticletitle \undefined \def \showarticletitle #1{#1}   \fi
\ifx \showURL      \undefined \def \showURL       {\relax}        \fi
\providecommand\bibfield[2]{#2}
\providecommand\bibinfo[2]{#2}
\providecommand\natexlab[1]{#1}
\providecommand\showeprint[2][]{arXiv:#2}

\bibitem[\protect\citeauthoryear{Amodei, Ananthanarayanan, Anubhai, Bai,
  Battenberg, Case, Casper, Catanzaro, Cheng, Chen, et~al\mbox{.}}{Amodei
  et~al\mbox{.}}{2016}]%
        {amodei2016deep}
\bibfield{author}{\bibinfo{person}{Dario Amodei}, \bibinfo{person}{Sundaram
  Ananthanarayanan}, \bibinfo{person}{Rishita Anubhai},
  \bibinfo{person}{Jingliang Bai}, \bibinfo{person}{Eric Battenberg},
  \bibinfo{person}{Carl Case}, \bibinfo{person}{Jared Casper},
  \bibinfo{person}{Bryan Catanzaro}, \bibinfo{person}{Qiang Cheng},
  \bibinfo{person}{Guoliang Chen}, {et~al\mbox{.}}}
  \bibinfo{year}{2016}\natexlab{}.
\newblock \showarticletitle{Deep speech 2: End-to-end speech recognition in
  english and mandarin}. In \bibinfo{booktitle}{\emph{International Conference
  on Machine Learning}}. \bibinfo{pages}{173--182}.
\newblock


\bibitem[\protect\citeauthoryear{Bahdanau, Cho, and Bengio}{Bahdanau
  et~al\mbox{.}}{2014}]%
        {bahdanau2014neural}
\bibfield{author}{\bibinfo{person}{Dzmitry Bahdanau},
  \bibinfo{person}{Kyunghyun Cho}, {and} \bibinfo{person}{Yoshua Bengio}.}
  \bibinfo{year}{2014}\natexlab{}.
\newblock \showarticletitle{Neural machine translation by jointly learning to
  align and translate}.
\newblock \bibinfo{journal}{\emph{arXiv preprint arXiv:1409.0473}}
  (\bibinfo{year}{2014}).
\newblock


\bibitem[\protect\citeauthoryear{Cohen, Afshar, Tapson, and van Schaik}{Cohen
  et~al\mbox{.}}{2017}]%
        {emnist}
\bibfield{author}{\bibinfo{person}{Gregory Cohen}, \bibinfo{person}{Saeed
  Afshar}, \bibinfo{person}{Jonathan Tapson}, {and} \bibinfo{person}{Andr{\'e}
  van Schaik}.} \bibinfo{year}{2017}\natexlab{}.
\newblock \showarticletitle{EMNIST: an extension of MNIST to handwritten
  letters}.
\newblock \bibinfo{journal}{\emph{arXiv preprint arXiv:1702.05373}}
  (\bibinfo{year}{2017}).
\newblock


\bibitem[\protect\citeauthoryear{Cun, Cortes, and Burges}{Cun
  et~al\mbox{.}}{1998}]%
        {mnist}
\bibfield{author}{\bibinfo{person}{Y.~Le Cun}, \bibinfo{person}{C. Cortes},
  {and} \bibinfo{person}{C.~J. Burges}.} \bibinfo{year}{1998}\natexlab{}.
\newblock \bibinfo{title}{The mnist database of handwritten digits}.
\newblock
\newblock


\bibitem[\protect\citeauthoryear{Goodfellow, Shlens, and Szegedy}{Goodfellow
  et~al\mbox{.}}{2015}]%
        {goodfellow}
\bibfield{author}{\bibinfo{person}{I.~J. Goodfellow}, \bibinfo{person}{J.
  Shlens}, {and} \bibinfo{person}{C. Szegedy}.}
  \bibinfo{year}{2015}\natexlab{}.
\newblock \showarticletitle{Explaining and harnessing adversarial examples}. In
  \bibinfo{booktitle}{\emph{Proceedings of the 2015 International Conference on
  Learning Representations.}}
\newblock


\bibitem[\protect\citeauthoryear{Guo and Potkonjak}{Guo and Potkonjak}{2018}]%
        {guo2018watermarking}
\bibfield{author}{\bibinfo{person}{Jia Guo} {and} \bibinfo{person}{Miodrag
  Potkonjak}.} \bibinfo{year}{2018}\natexlab{}.
\newblock \showarticletitle{Watermarking deep neural networks for embedded
  systems}. In \bibinfo{booktitle}{\emph{Proceedings of the International
  Conference on Computer-Aided Design}}. ACM, \bibinfo{pages}{133}.
\newblock


\bibitem[\protect\citeauthoryear{Han, Pool, Tran, and Dally}{Han
  et~al\mbox{.}}{2015}]%
        {han2015learning}
\bibfield{author}{\bibinfo{person}{Song Han}, \bibinfo{person}{Jeff Pool},
  \bibinfo{person}{John Tran}, {and} \bibinfo{person}{William Dally}.}
  \bibinfo{year}{2015}\natexlab{}.
\newblock \showarticletitle{Learning both weights and connections for efficient
  neural network}. In \bibinfo{booktitle}{\emph{Advances in neural information
  processing systems}}. \bibinfo{pages}{1135--1143}.
\newblock


\bibitem[\protect\citeauthoryear{Hannun, Case, Casper, Catanzaro, Diamos,
  Elsen, Prenger, Satheesh, Sengupta, Coates, et~al\mbox{.}}{Hannun
  et~al\mbox{.}}{2014}]%
        {hannun2014deep}
\bibfield{author}{\bibinfo{person}{Awni Hannun}, \bibinfo{person}{Carl Case},
  \bibinfo{person}{Jared Casper}, \bibinfo{person}{Bryan Catanzaro},
  \bibinfo{person}{Greg Diamos}, \bibinfo{person}{Erich Elsen},
  \bibinfo{person}{Ryan Prenger}, \bibinfo{person}{Sanjeev Satheesh},
  \bibinfo{person}{Shubho Sengupta}, \bibinfo{person}{Adam Coates},
  {et~al\mbox{.}}} \bibinfo{year}{2014}\natexlab{}.
\newblock \showarticletitle{Deep speech: Scaling up end-to-end speech
  recognition}.
\newblock \bibinfo{journal}{\emph{arXiv preprint arXiv:1412.5567}}
  (\bibinfo{year}{2014}).
\newblock


\bibitem[\protect\citeauthoryear{He, Zhang, Ren, and Sun}{He
  et~al\mbox{.}}{2016}]%
        {resnet}
\bibfield{author}{\bibinfo{person}{Kaiming He}, \bibinfo{person}{Xiangyu
  Zhang}, \bibinfo{person}{Shaoqing Ren}, {and} \bibinfo{person}{Jian Sun}.}
  \bibinfo{year}{2016}\natexlab{}.
\newblock \showarticletitle{Deep residual learning for image recognition}. In
  \bibinfo{booktitle}{\emph{Proceedings of the IEEE conference on computer
  vision and pattern recognition}}. \bibinfo{pages}{770--778}.
\newblock


\bibitem[\protect\citeauthoryear{Hinton and Salakhutdinov}{Hinton and
  Salakhutdinov}{2006}]%
        {ae}
\bibfield{author}{\bibinfo{person}{Geoffrey Hinton} {and}
  \bibinfo{person}{Ruslan Salakhutdinov}.} \bibinfo{year}{2006}\natexlab{}.
\newblock \showarticletitle{Reducing the Dimensionality of Data with Neural
  Networks}.
\newblock \bibinfo{journal}{\emph{Science}} \bibinfo{volume}{313},
  \bibinfo{number}{5786} (\bibinfo{year}{2006}), \bibinfo{pages}{504 -- 507}.
\newblock


\bibitem[\protect\citeauthoryear{Krizhevsky}{Krizhevsky}{2009}]%
        {cifar10}
\bibfield{author}{\bibinfo{person}{Alex Krizhevsky}.}
  \bibinfo{year}{2009}\natexlab{}.
\newblock \bibinfo{title}{Learning Multiple Layers of Features from Tiny
  Images}.
\newblock
\newblock


\bibitem[\protect\citeauthoryear{Merrer, Perez, and Tr{\'e}dan}{Merrer
  et~al\mbox{.}}{2017}]%
        {merrer2017adversarial}
\bibfield{author}{\bibinfo{person}{Erwan~Le Merrer}, \bibinfo{person}{Patrick
  Perez}, {and} \bibinfo{person}{Gilles Tr{\'e}dan}.}
  \bibinfo{year}{2017}\natexlab{}.
\newblock \showarticletitle{Adversarial frontier stitching for remote neural
  network watermarking}.
\newblock \bibinfo{journal}{\emph{arXiv preprint arXiv:1711.01894}}
  (\bibinfo{year}{2017}).
\newblock


\bibitem[\protect\citeauthoryear{Miyato, Maeda, Ishii, and Koyama}{Miyato
  et~al\mbox{.}}{2018}]%
        {vat}
\bibfield{author}{\bibinfo{person}{T. Miyato}, \bibinfo{person}{S. Maeda},
  \bibinfo{person}{S. Ishii}, {and} \bibinfo{person}{M. Koyama}.}
  \bibinfo{year}{2018}\natexlab{}.
\newblock \showarticletitle{Virtual Adversarial Training: A Regularization
  Method for Supervised and Semi-Supervised Learning}.
\newblock \bibinfo{journal}{\emph{IEEE Transactions on Pattern Analysis and
  Machine Intelligence}} (\bibinfo{year}{2018}), \bibinfo{pages}{1--1}.
\newblock
\showISSN{0162-8828}
\urldef\tempurl%
\url{https://doi.org/10.1109/TPAMI.2018.2858821}
\showDOI{\tempurl}


\bibitem[\protect\citeauthoryear{Rouhani, Chen, and Koushanfar}{Rouhani
  et~al\mbox{.}}{2018}]%
        {rouhani2018deepsigns}
\bibfield{author}{\bibinfo{person}{Bita~Darvish Rouhani},
  \bibinfo{person}{Huili Chen}, {and} \bibinfo{person}{Farinaz Koushanfar}.}
  \bibinfo{year}{2018}\natexlab{}.
\newblock \showarticletitle{Deepsigns: A generic watermarking framework for ip
  protection of deep learning models}.
\newblock \bibinfo{journal}{\emph{arXiv preprint arXiv:1804.00750}}
  (\bibinfo{year}{2018}).
\newblock


\bibitem[\protect\citeauthoryear{Sebastian, Johannes, Jan, Marc, and
  Christian}{Sebastian et~al\mbox{.}}{2013}]%
        {gtsrb}
\bibfield{author}{\bibinfo{person}{Houben Sebastian},
  \bibinfo{person}{Stallkamp Johannes}, \bibinfo{person}{Salmen Jan},
  \bibinfo{person}{Schlipsing Marc}, {and} \bibinfo{person}{Igel Christian}.}
  \bibinfo{year}{2013}\natexlab{}.
\newblock \showarticletitle{Detection of Traffic Signs in Real-World Images:
  The {G}erman {T}raffic {S}ign {D}etection {B}enchmark}. In
  \bibinfo{booktitle}{\emph{International Joint Conference on Neural
  Networks}}.
\newblock


\bibitem[\protect\citeauthoryear{Simonyan and Zisserman}{Simonyan and
  Zisserman}{2014}]%
        {simonyan2014very}
\bibfield{author}{\bibinfo{person}{Karen Simonyan} {and}
  \bibinfo{person}{Andrew Zisserman}.} \bibinfo{year}{2014}\natexlab{}.
\newblock \showarticletitle{Very deep convolutional networks for large-scale
  image recognition}.
\newblock \bibinfo{journal}{\emph{arXiv preprint arXiv:1409.1556}}
  (\bibinfo{year}{2014}).
\newblock


\bibitem[\protect\citeauthoryear{Szegedy, Zaremba, Sutskever, Bruna, Erhan,
  Goodfellow, and Fergus}{Szegedy et~al\mbox{.}}{2013}]%
        {szegedy2013intriguing}
\bibfield{author}{\bibinfo{person}{Christian Szegedy},
  \bibinfo{person}{Wojciech Zaremba}, \bibinfo{person}{Ilya Sutskever},
  \bibinfo{person}{Joan Bruna}, \bibinfo{person}{Dumitru Erhan},
  \bibinfo{person}{Ian Goodfellow}, {and} \bibinfo{person}{Rob Fergus}.}
  \bibinfo{year}{2013}\natexlab{}.
\newblock \showarticletitle{Intriguing properties of neural networks}.
\newblock \bibinfo{journal}{\emph{arXiv preprint arXiv:1312.6199}}
  (\bibinfo{year}{2013}).
\newblock


\bibitem[\protect\citeauthoryear{Uchida, Nagai, Sakazawa, and Satoh}{Uchida
  et~al\mbox{.}}{2017}]%
        {uchida2017embedding}
\bibfield{author}{\bibinfo{person}{Yusuke Uchida}, \bibinfo{person}{Yuki
  Nagai}, \bibinfo{person}{Shigeyuki Sakazawa}, {and}
  \bibinfo{person}{Shin'ichi Satoh}.} \bibinfo{year}{2017}\natexlab{}.
\newblock \showarticletitle{Embedding watermarks into deep neural networks}. In
  \bibinfo{booktitle}{\emph{Proceedings of the 2017 ACM on International
  Conference on Multimedia Retrieval}}. ACM, \bibinfo{pages}{269--277}.
\newblock


\bibitem[\protect\citeauthoryear{Yosinski, Clune, Bengio, and Lipson}{Yosinski
  et~al\mbox{.}}{2014}]%
        {yosinski2014transferable}
\bibfield{author}{\bibinfo{person}{Jason Yosinski}, \bibinfo{person}{Jeff
  Clune}, \bibinfo{person}{Yoshua Bengio}, {and} \bibinfo{person}{Hod Lipson}.}
  \bibinfo{year}{2014}\natexlab{}.
\newblock \showarticletitle{How transferable are features in deep neural
  networks?}. In \bibinfo{booktitle}{\emph{Advances in neural information
  processing systems}}. \bibinfo{pages}{3320--3328}.
\newblock


\bibitem[\protect\citeauthoryear{Yossi, Carsten, Moustapha, Benny, and
  Joseph}{Yossi et~al\mbox{.}}{2018}]%
        {adi2018turning}
\bibfield{author}{\bibinfo{person}{Adi Yossi}, \bibinfo{person}{Baum Carsten},
  \bibinfo{person}{Cisse Moustapha}, \bibinfo{person}{Pinkas Benny}, {and}
  \bibinfo{person}{Keshet Joseph}.} \bibinfo{year}{2018}\natexlab{}.
\newblock \showarticletitle{Turning Your Weakness Into a Strength: Watermarking
  Deep Neural Networks by Backdooring}. In \bibinfo{booktitle}{\emph{27th
  {USENIX} Security Symposium ({USENIX} Security 18)}}.
  \bibinfo{publisher}{{USENIX} Association}.
\newblock


\bibitem[\protect\citeauthoryear{Zhang, Gu, Jang, Wu, Stoecklin, Huang, and
  Molloy}{Zhang et~al\mbox{.}}{2018}]%
        {zhang2018protecting}
\bibfield{author}{\bibinfo{person}{Jialong Zhang}, \bibinfo{person}{Zhongshu
  Gu}, \bibinfo{person}{Jiyong Jang}, \bibinfo{person}{Hui Wu},
  \bibinfo{person}{Marc~Ph Stoecklin}, \bibinfo{person}{Heqing Huang}, {and}
  \bibinfo{person}{Ian Molloy}.} \bibinfo{year}{2018}\natexlab{}.
\newblock \showarticletitle{Protecting Intellectual Property of Deep Neural
  Networks with Watermarking}. In \bibinfo{booktitle}{\emph{Proceedings of the
  2018 on Asia Conference on Computer and Communications Security}}. ACM,
  \bibinfo{pages}{159--172}.
\newblock


\end{thebibliography}

\clearpage

\appendix

\section{Architecture of the autoencoder and model hyperparameters}\label{ap:arch}

\tb{aec-arch} shows the architecture of our convolutional autoencoder and \tb{param} summarizes the hyperparameters used to train ResNet and autoencoder.

\begin{table}[!h]
  \begin{center}
    \caption{Architecture of autoencoder. The layertype of ``conv'', ``deconv'' and ``bn'' means convolution, deconvolution and batch normalization respectively. Relu and sigmoid means activation function.}
    \label{tb:aec-arch}
    \scalebox{0.8}{
    \begin{tabular}{c|cc} \hline
      layer type & kernel size & stride \\ \hline
      conv,bn(relu)  & $3 \times 3 \times 16$ & 2 \\
      conv,bn(relu)  & $3 \times 3 \times 32$ & 2 \\
      conv,bn(relu)  & $3 \times 3 \times 64$ & 2 \\
      deconv,bn(relu) & $3 \times 3 \times 32$ & 2 \\
      deconv,bn(relu) & $3 \times 3 \times 16$ & 2 \\
      deconv,bn(sigmoid) & $3 \times 3 \times 3$ & 2\\ \hline
    \end{tabular}
    }
  \end{center}
\end{table}

\begin{table}[!h]
  \begin{center}
    \caption{Model parameters.}
    \label{tb:param}
    \scalebox{0.8}{
    \begin{tabular}{c|cc} \hline
      parameter & resnet & autoencoder \\ \hline
      optimization & Momentum SGD & Adam \\
      epoch & $100$ & $400$ \\
      mini batch size & $100$ & $100$ \\
      key mini batch size & $4$ & - \\
      learning rate & 0.1($ \times 0.1$ when epoch is $40,60$) & 0.001 \\ \hline
    \end{tabular}
    }
  \end{center}
\end{table}

\section{Assignments of samples}\label{ap:assignment}

\tb{num_samples} shows (1) the number of samples that the model owner used to train classification model, (2) the number of key samples that the model owner embedded into the model and (3) sample size that the unauthorized service provider utilized to invalidate watermark, in each dataset. The total numbers of training samples of each dataset are 60,000 (MNIST), 39,209 (GTSRB) and 50,000 (CIFAR10, CIFAR100).
We suppose the unauthorized service provider can use 1\% (MNIST), 5\% (GTSRB) and 10\% (CIFAR10, CIFAR100) of total training samples for invalidation.
Since MNIST and GTSRB are relatively easy classification task, if the unauthorized service provider has a sufficiently large number of samples (as ten \% in CIFAR10 and CIFAR100), the unauthorized service provider can train a model that can achieve a sufficiently high test accuracy. In such a situation, the unauthorized service provider does not have the motivation to use licensed models without permission. For this reason, we assume the unauthorized user has a smaller number of samples for invalidation for MNIST and GTSRB than CIFAR10 and CIFAR100.

The number of key samples embedded into the model is different for each method for the following reasons.
Embedding by \cite{zhang2018protecting}-content requires that the model well generalizes the symbols or logotypes to achieve good AUC. It is thus desirable to embed a large number of key samples.
Watermarking using \cite{merrer2017adversarial}-AFS and \cite{rouhani2018deepsigns}-DS take advantage of model overfitting to specific key samples. So, we embedded 30 key samples into the model following the recommendation of by \cite{merrer2017adversarial} and \cite{rouhani2018deepsigns}.


\begin{table}[h]
  \begin{center}
    \caption{The number of samples used for training, and watermark, and model invalidation. Each column shows (1) the number of samples that the model owner used to train the classification model (training samples), (2) the number of samples that the model owner embedded into the model (key samples), and (3) the number of samples that the unauthorized service provider utilized to detect key samples in each dataset.}
    \label{tb:num_samples}
    \scalebox{0.8}{
      \begin{tabular}{cc|cccc} \hline
        \multicolumn{2}{c|}{dataset} & MNIST & GTSRB & CIFAR10 & CIFAR100 \\ \hline
        \multicolumn{2}{c|}{\#training samples} & 59,400 & 37,248 & 45,000 & 45,000 \\
        & \cite{zhang2018protecting}-content & 53,548 & 37,046 & 40,478 & 44,558 \\
        & \cite{zhang2018protecting}-unrelated & 2,400 & 6,742 & 6,742 & 6,742 \\
        \#key samples for: & \cite{zhang2018protecting}-noise & 53,548 & 37,046 & 40,478 & 44,558 \\
        & \cite{merrer2017adversarial}-AFS & 30 & 30 & 30 & 30 \\
        & \cite{rouhani2018deepsigns}-DS & 30 & 30 & 30 & 30 \\
        \multicolumn{2}{c|}{\#samples used for key sample detection} & 600 & 1,961 & 5,000 & 5,000 \\ \hline
      \end{tabular}
      }
  \end{center}
\end{table}

\end{document}